\pdfoutput = 1
\documentclass[twocolumn]{svjour3}          
\smartqed  
\usepackage{graphicx}

\journalname{}
\usepackage[misc]{ifsym}
\usepackage[usenames]{color}
\usepackage{subfigure}
\usepackage{cite}
\usepackage{enumitem}
\usepackage{mathrsfs}
\usepackage{amsfonts}
\usepackage{amsmath,mathtools}
\usepackage{algorithm}
\usepackage{algpseudocode}
\usepackage{multirow}
\usepackage[OT1]{fontenc}
\usepackage[utf8]{inputenc}
\usepackage[english]{babel}
\usepackage{hyperref}

\newcommand{\red}{\color{black}}

\newtheorem{defn}{Definition}

\newcommand{\mR}{\mathbb{R}}
\def\N{{\mathcal N}}
\def\T{{\mathcal T}}

\mathtoolsset{showonlyrefs=true}

\begin{document}

\title{The Graph-Based Behavior-Aware Recommendation \\ for Interactive News}

\author{Mingyuan Ma \textsuperscript{1}  \and Sen Na \textsuperscript{2} \and Hongyu Wang \textsuperscript{1,3} \and Congzhou Chen \textsuperscript{1} \and Jin Xu \textsuperscript{1}}
\authorrunning{Ma et al.}  

\institute{
\Letter ~Jin Xu\\
\email{jxu@pku.edu.cn}\\
\Letter ~Hongyu Wang\\
\email{why5126@pku.edu.cn}\\
\at
 {1} School of Electronics Engineering and Computer Science, Peking University
 \at
 {2} Department of Statistics, University of Chicago
 \at
 {3} National Computer Network Emergency Response Technical Team/Coordination Center of China\\
}

\date{Received: date / Accepted: date}

\maketitle

\begin{abstract}

Interactive news recommendation has been launched and attracted much attention recently. In this scenario, user's behavior evolves from single click behavior to multiple behaviors including like, comment, share etc. However, most of the existing methods still use single click behavior as the unique criterion of judging user's preferences. Further, although heterogeneous graphs have been applied in different areas, a proper way to construct a heterogeneous graph for interactive news data with an appropriate learning mechanism on it is still desired. To address the above concerns, we propose a graph-based behavior-aware network, which simultaneously considers six different types of behaviors as well as user's demand on the news diversity. We have three main steps. First, we build an interaction behavior graph for multi-level and multi-category data. Second, we apply DeepWalk on the behavior graph to obtain entity semantics, then build a graph-based convolutional neural network called G-CNN to learn news representations, and an attention-based LSTM to learn behavior sequence representations. Third, we introduce core and coritivity features for the behavior graph, which measure the concentration degree of user's interests. These features affect the trade-off between accuracy and diversity of our personalized recommendation system. Taking these features into account, our system finally achieves recommending news to different users at their different levels of concentration degrees.

\end{abstract}

\section{Introduction}\label{intro}

\begin{figure}[!tp]
	\centering
	\subfigure[news page]{\label{fig:1a}
		\includegraphics[width=1.3in]{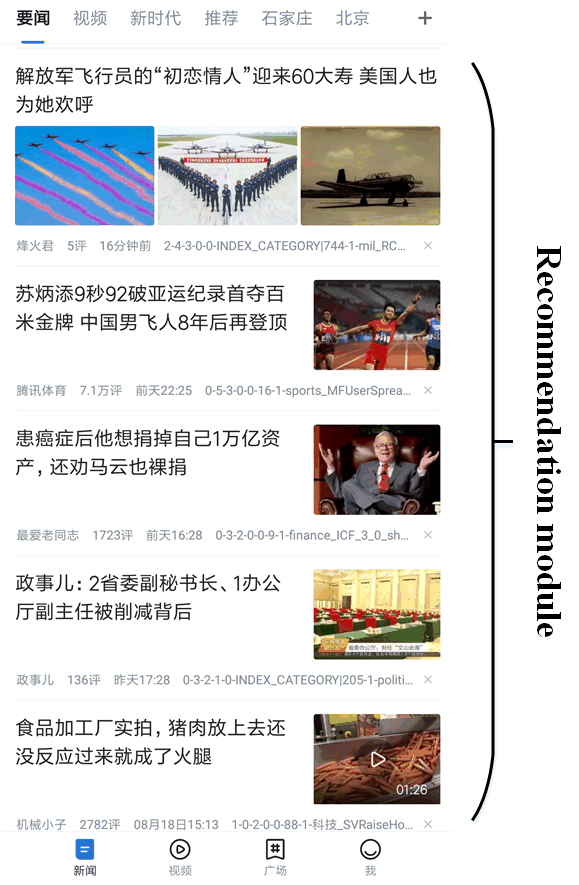}}
	\hspace{0.18in}
	\subfigure[interactive news page]{\label{fig:1b}
		\includegraphics[width=1.4in]{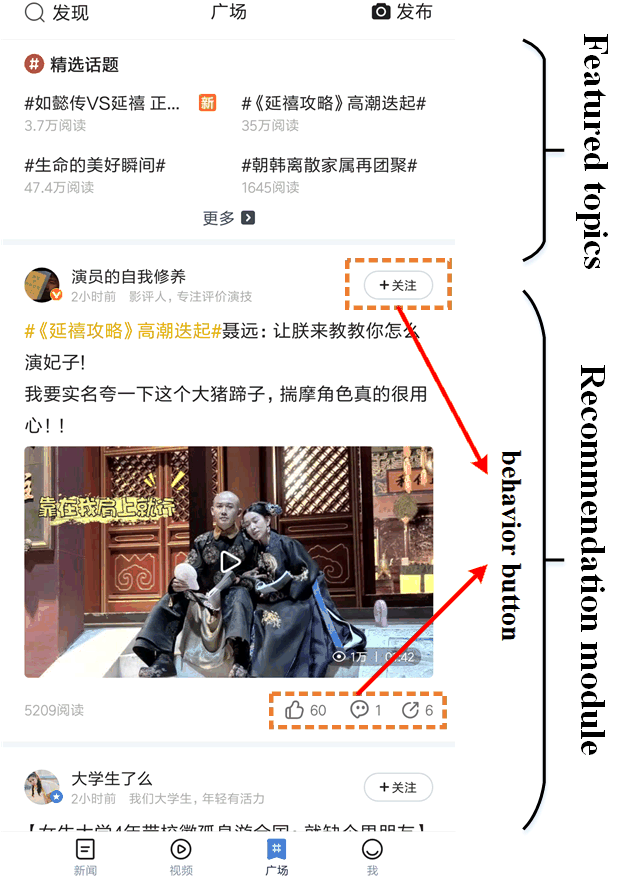}}
	\caption{Tencent's news page and interactive news page on smartphones (in Chinese characters).}
	\label{fig:1}
\end{figure}

Because of high accessibility, electronic media, such as news apps and social networking apps on a smartphone, are challenging traditional paper media and becoming the primary way for people to access new information. According to statistics, the number of monthly active users of Tencent News App, one of the most popular apps in China, had surpassed 280 million in the first half of 2018 and kept increasing rapidly between 2018 and 2021. To improve users experience, many news content providers have launched a series of interactive news pages. Figure \ref{fig:1} is the mockup of Tencent News App. On the news page, users usually glance over top news or click an item of news to read thoroughly if they think it is interesting, while on the interactive page, users can view hot news post by other users via an interactive way: publish own viewpoints, share the news, and follow the bloggers etc. Therefore, in the interactive news scenario, users tend to have diverse behaviors, which contain more implicit evidence on their interests.

Traditionally, researchers have studied recommendation systems under different setups. We may have collaborative filtering (CF) recommendation \cite{Breese1998Empirical, Sarwar2001Item}, content-based (CB) recommendation \cite{Balabanovic1997Fab, Basu1998Recommendation, Pazzani2007Content}, and hybrid recommendation \cite{Burke2002Hybrid, Kardan2013novel}. The technical tools include latent dirichlet allocation \cite{Blei2003Latent}, Bayesian matrix factorization \cite{Salakhutdinov2008Bayesian}, (inductive) matrix completion \cite{Shin2015Tumblr, Na2020Semiparametric}, and $k$-nearest neighbor (kNN) method \cite{Bell2007Improved}. We refer to \cite{Ricci2015Recommender} for a brief survey on recommendation systems. Over the past decade, the deep learning methods have shown ground-breaking performance in a variety of domains ranging from image recognition \cite{Simonyan2015Very, He2016Deep}, speech recognition \cite{Zhang2017Very}, to machine translation \cite{Papineni2002BLEU, Cho2014Learning, Bahdanau2015Neural, Nakayama2017Zero}. Recently, researchers have also applied convolutional neural network (CNN) and recurrent neural network (RNN) in many recommendation scenarios and achieved good performance. For example, a recent work \cite{Okura2017Embedding} proposed a RNN architecture to generate users representations using browsing histories as input sequences. The authors showed in experiments on recommending Yahoo news that the click-through rate (CTR) is enhanced by 23\% using their deep learning framework. In addition, other advanced deep-learning-based recommendation systems, such as Collaborative Deep Learning \cite{Wang2015Collaborative}, DeepFM \cite{Guo2017DeepFM}, YouTubeNet \cite{Covington2016Deep}, Neural Collaborative Filtering \cite{He2017Neural}, NFM \cite{He2017Neurala}, AFM \cite{Xiao2017Attentional}, DAN \cite{Zhu2019DAN}, and KGAT \cite{Wang2019Kgat}, have also been reported. Our work is closely related to this line of literature and detailed comparisons will be introduced later.

Although research on news recommendation has been investigated from different aspects, most of existing methods still use single click behavior as the unique criterion of judging user's preferences. This may be acceptable for recommending traditional news since the click behavior is the only feedback from users. However, for interactive news, it will result in a biased prediction due to the neglect of other behaviors. For example, if a user \textit{follows} many bloggers who post the same topic of news, it strongly suggests that we should recommend news in this topic to this user, and such information will be lost if one only focuses on the click behavior. Further, in a more realistic case, the concentration degree of user's interests varies from user to user. Some users may be interested in multiple news categories, while some may concentrate on a specific category of news. Therefore, each user has own demand on the news diversity extent, and such differences between users are more evident in reading interactive news. Intuitively, a user with high concentration degree tends to generate lots of similar behaviors for a certain type of news, while a user with low concentration degree tends to have more diverse behaviors. Instead of using a global feature to control the whole system diversity \cite{Javari2015probabilistic}, a more appropriate way is to extract a user-dependent concentration feature and make it involved when building the system. Moreover, we note that users and news together with multiple behaviors, categories, and topics naturally constitute a weighted heterogeneous graph. This graph almost contains all useful information for recommendation. However, it is difficult to model and extract those heterogeneous and complex information. Targeting the above limitations, the following question arises naturally:

\emph{How to build a learning framework on heterogeneous graph-structured data that can deal with multiple behaviors and, simultaneously, recommend news based on the user's interests concentration degree.}

In this work, we answer the above question by our proposed model called graph-based behavior-aware network (GBAN). GBAN leverages tools from broad learning \cite{Zhang2018Broad}, graph theory \cite{Wu2014Coritivity}, and domain-specific knowledge graph \cite{Xiong2017Explicit} to obtain suitable users and news representations, and then it feeds the learned representations into a two-layer fully-connected neural network to perform a six-classification task. Here, the six possible behaviors are: \textit{unclick}, \textit{click}, \textit{like}, \textit{follow}, \textit{comment}, and \textit{share}. We point out that ``unclick'' behavior would occur when the other five behaviors could occur but did not. GBAN inputs a sequence of user's behaviors and a piece of candidate news, and outputs the probability of each behavior that the user may perform on the candidate news. We emphasize that designing a system under such setup is more reasonable for interactive recommendation than a biclassification task using the click-through rate (CTR) as the performance metric.

\vskip5pt
\noindent \textbf{GBAN structure.} We briefly introduce the structure of GBAN. Details of each step are demonstrated in the next two sections. Given the data including users, news, behavior sequences of users, and categories, topics, and tags of news, GBAN consists of four steps:
\begin{enumerate}[label=(\alph*),topsep=0pt]
\setlength\itemsep{0.0em}

\item We build an interaction behavior graph. Each node represents either a user, a piece of news, a category, a tag, or a topic, and whether two nodes are connected by an edge is determined by whether they have realistic connections. For example, the edge between a user node and a news node implies that one of five (positive) behaviors is triggered in the history. Further, we assign this edge a weight according to the behavior type, which increases proportionally from click to share. The details of assigning weights for all types of edges are described in Section \ref{sec:3}.

\item Based on the behavior graph, we perform two tasks: calculate \textit{core} and \textit{coritivity} for subgraphs induced by each user as his/her concentration feature, and learn a vector representation for each node. Here, core and coritivity are concepts from graph theory \cite{Jin1999CORE}, which quantitatively measure the importance of the nodes and provide useful insights for user's concentration degree. Detailed explanations and algorithms are presented in Sections \ref{sec:2} and \ref{sec:3}. The vector representations are learned by DeepWalk \cite{Perozzi2014Deepwalk}.

\item Using vector representations, we design a graph-based convolutional network called G-CNN to learn news representations. Further, we design an attention-based LSTM to learn user's behavior sequence representations. In particular, given a specific user and a specific candidate news, LSTM takes the concatenated news representations from G-CNN as inputs, with attention weights being determined by user node representations and candidate news node representations jointly. Here, the concatenation of news representations is ordered according to the temporal behavior sequence. Finally, the behavior sequence representations together with concentration feature (from Step (b)) constitute the user representations.

\item We combine user representations with candidate news representations, feed them into two fully-connected layers to perform a six-classification task.
	
\end{enumerate}
The above four steps are the main workflow of GBAN.

To wrap up, interactive news recommendation is a new scenario and there are very limited works targeting this problem so far. The main challenges lie in the analysis of multiple behaviors feedback from users and the personalized demand on the news diversity. Our proposed GBAN method resolves these challenges and has following contributions:
\begin{enumerate}[label=(\alph*),topsep=0pt]
\setlength\itemsep{0.0em}

\item We structure the interactive news data as an interaction behavior graph, which is a weighted heterogeneous graph. To our knowledge, this is the first paper that clearly illustrates how to compress the interactive news data containing multiple behaviors and complex users and news profiles by a heterogeneous graph. A clear advantage of studying such graph is that the representations of different types of nodes are learned in the same space. Thus, we need not involve a complex space transformation when we consider different types of nodes.

\item The proposed method GBAN is able to recommend interactive news by considering six types of behaviors. GBAN applies a G-CNN to learn news representations, and an attention-based LSTM to learn behavior sequence representations. Our architecture is simpler than the state-of-the-art model DAN \cite{Zhu2019DAN}, but achieves better performance. In addition, our attention mechanism for LSTM depends on the specific candidate news, which enables LSTM to learn behavior sequence representations with pertinence.

\item We propose the concept of user's interests concentration degree, which characterizes his/her demand on the news diversity. To quantitatively measure the concentration degree, our paper proposes a simple resolution: core and coritivity of the subgraph induced by the user. By including the concentration feature, GBAN can adaptively recommend news according to different diversity levels. To our knowledge, this paper is the first endeavor to incorporate this concentration feature into the recommendation.

\item We apply GBAN on two datasets. One is the dataset from Tencent News App and one is the MovieLens dataset. We show the superiority of GBAN through comprehensive comparisons with other state-of-the-art methods and its own variants.

\end{enumerate}

\vskip 4pt
\noindent{\bf Related work.} Our work is related to a growing literature on deep learning-based recommendation system. To highlight a few of them, Okura et al. \cite{Okura2017Embedding} adopted a RNN to generate user representations using browsing histories as input sequences. DeepFM \cite{Guo2017DeepFM} studied the feature interactions behind user behaviors via a novel neural network model, which combines the power of factorization machines and deep learning for feature learning. YouTubeNet \cite{Covington2016Deep} applied the classic two-stage information retrieval dichotomy for video recommendation, achieved by deep candidate generation model and deep ranking model, respectively. Collaborative Deep Learning \cite{Wang2015Collaborative}, a hierarchical Bayesian model, addressed the sparsity issue in ratings, which outperforms conventional CF-based methods significantly. Neural CF \cite{He2017Neural} focused on generalizing matrix factorization and replacing the inner product between the latent features of users and items by multi-layer perceptron. Bobadilla et al. \cite{Bobadilla2020Classification} designed a classification-based deep learning CF approach that operates on the ratings data. The learning process is based on two binary sources: relevant/non-relevant vote and voted/non-voted item. Hurtado et al. \cite{Hurtado2020Collaborative} provided a method that combines diverse machine learning algorithms to make recommendation to large homogeneous user groups. The homogeneous groups are detected by performing clustering on hidden factors of users, and a virtual user is obtained for each group by performing hidden factors aggregation. All aforementioned work proposed advanced deep-learning-based recommendation systems. However, most of them are less concerned about several issues including the contextual connection and the demand on news diversity, which make them not suitable for interactive news recommendation.

To address those issues, people tried to build personalized recommendation systems by digging out other auxiliary features and/or taking advantage of representation learning on different graphs with the attention module. For example, Lian et al. \cite{Lian2018Towards} developed a deep fusion model. There are two critical components: an inception module and an attention mechanism. The former is composed of multi-layer networks and learns various levels of feature interaction, while the latter merges latent representations learned from different channels in a customized fashion. Subsequently, DKN \cite{Wang2018DKN} built a knowledge graph to extract latent knowledge-level connections among news. Within DKN, a multi-channel CNN is adopted to fuse semantic-level and knowledge-level representations of news. In addition, an attention module is adopted to address the user's diverse interests by aggregating the user's history with respect to the candidate news. However, both methods have flaws when they are applied on interactive news scenario. The former method \cite{Lian2018Towards} does not make use of the graph structure of data and the diversity in their study is limit to the behavior patterns, instead of the diversity of user's interests concentration degree. The latter method \cite{Wang2018DKN} separates the features of users and news from each other so that some implicit connections between users features and news features are neglected. Although the knowledge graph in DKN can aggregate the entity-based information, it fails to capture the news sequential information. Recently, DAN \cite{Zhu2019DAN} addressed the drawbacks of DKN by adopting an attention-based RNN to capture the hidden time series features, and achieved a better performance than DKN. Our paper competes with DAN thoroughly, and shows that GBAN outperforms DAN further with a simpler architecture. On the other hand, in order to design a user-oriented product, Alonso-Virg\'os et al. \cite{AlonsoVirgos2019Analyzing, AlonsoVirgos2020Test} analyzed the compliance and usability in developing websites. The authors used some statistical tools to study the behavior of web developers, and further to determine how to recommend the usability guidance to web developers in a more effective way. Comparing with this sequence of methods, the interactive news recommendation is more challenging. We have to exploit the data structure to distinguish multiple behaviors (i.e. six possible behaviors) of each user and also to estimate the concentration feature. This motivates us to build a heterogeneous behavior graph and leverage tools from graph theory.

In addition, it is worth mentioning that some recent work studied recommendation with multiple feedback types. Wan et al. \cite{Wan2018Item} observed that feedback signals exhibit monotonic dependency structures, i.e., any explicit signal necessarily implies the presence of implicit signal (a ``review" action implies a ``purchase" action, which implies a ``click" action, etc.). Gao et al. \cite{Gao2019Neural} designed Neural Multi-Task Recommendation (NMTR) method for learning multi-behavior data. Both methods assumed that the behavior types follow a sequential relationship, which is unrealistic for interactive news scenario. The multiple behaviors in our paper don't follow the particular ordinal relationship. For example, ``follow" and ``share" have no implication between them. Thus, a novel approach to deal with multiple behaviors especially for interactive news is still desired.

\vskip 5pt
\noindent{\bf Structure of the paper.} In Section \ref{sec:2}, we introduce preliminaries including core and coritivity concepts, graph representation learning, and also formulate our recommendation problem. In Section \ref{sec:3}, we delve into the details of our GBAN method. We then conduct extensive experiments and compare with different advanced models on real data in Section \ref{sec:4}, and present conclusions and future work in Section \ref{sec:5}.

\section{Preliminary and Problem Formulation}\label{sec:2}

In this section, we introduce core and coritivity concepts and graph representation learning. Then we formulate the recommendation problem considered in this paper. Throughout the presentation, the terminologies of node and entity are exchangeable.

\subsection{Core and Coritivity}

In a graph network, there are always some entities that are locating at important positions and playing crucial roles. Removing these entities may lead the graph to enter a decentralized state. We call these entities cores of the graph network.

\begin{figure*}[!t]
\centering
\includegraphics[scale=0.32]{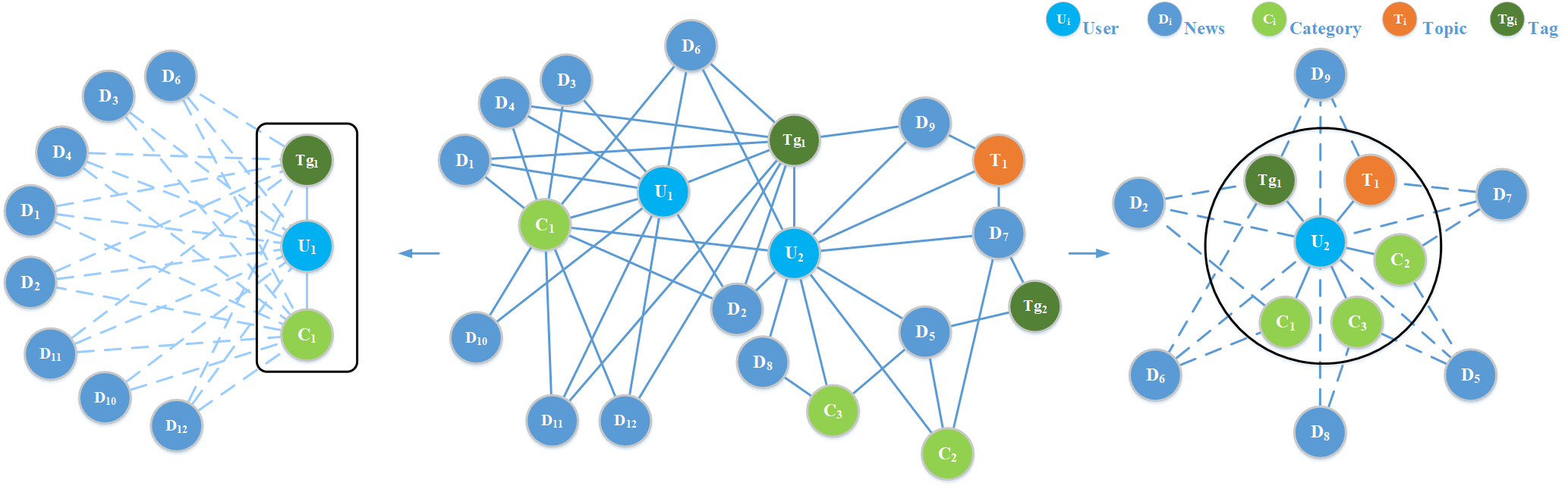}
\caption{The illustration of different concentration degrees of users' interests. The middle panel is the interaction behavior graph; the left panel is the subgraph induced by the user $U_1$; and the right panel is the subgraph induced by the user $U_2$.}
\label{fig:c}
\end{figure*}

In the interactive news scenario, we first construct a weighted heterogeneous graph, as shown in the middle panel of Figure \ref{fig:c}. In such a graph, each node represents either a user, a piece of news, a category, a topic, or a tag, and the edges are determined by the triggered behaviors. We then extract two subgraphs induced by users $U_1$ and $U_2$ respectively, by including only the specific user node and its all adjacent nodes, as shown in the left and the right panels of Figure \ref{fig:c}. Within each user subgraph, we see from the figure that if we remove the nodes in the box (or the circle), the subgraph will enter a decentralized state. Thus, the nodes in the box (or the circle) play an important role and are cores of the subgraph. Comparing the number of nodes in the box and circle and their dispersion extent, we can intuitively conclude that two users own different demand on the news diversity. In particular, $U_1$ is interested in reading a specific category of news, while $U_2$ is interested in multiple news categories.

To rigorously study cores and their effects, the paper \cite{Jin1999CORE} formally proposed the concepts of core and coritivity, which quantitatively measure the importance of a set of nodes by the number of connected components showing up after deleting the nodes and their incident edges. We present the definitions as follows.

\begin{defn}[Coritivity]\label{def:1}

Given an undirected connected graph $G$, we define its coritivity $h(G)$ as:
\begin{equation}\label{eq:1}
h(G)=\max\{\omega(G-S)-|S|: S \in C(G)\},
\end{equation}
where $C(G)$ denotes the set of all vertex cuts of $G$, and $\omega(G - S)$ is the number of connected components of $G - S$. For $S \in C(G)$, $G-S$ denotes the graph obtained by deleting the node set $S$ together with all edges incident with any nodes in $S$. $|S|$ denotes the number of nodes in $S$.
\end{defn}

\begin{defn}[Core]\label{def:2}
The set $S$ is called a core of graph $G$, if $S \in C(G)$ and satisfies
\begin{equation}\label{eq:2}
h(G)=\omega(G-S)-|S|.
\end{equation}
\end{defn}

According to Definitions \ref{def:1} and \ref{def:2}, it's useful to note that each graph has a unique coritivity value but may have many different cores. Each core is a cut set that satisfies \eqref{eq:2}. We choose the core such that $|S|$ is as small as possible. Our work uses core and coritivity pair, $(|S|, h(G))$, of the subgraph induced by each user to measure his/her concentration degree. It's easy to see that the larger $h(G)$ or $|S|$ implies the more connected components or important nodes, further implies the lower concentration degree and the larger demand on news diversity. In principle, a user with high concentration degree tends to generate a lot of similar behaviors for a certain type of news. On the contrary, a user with low concentration degree tends to generate more diverse behaviors.

Since the coritivity indicates the importance of core nodes, given a graph, the most fundamental problem is to design an algorithm to calculate the coritivity. Zhang et al.  \cite{Zhang2016Structural} proved that exactly computing the coritivity is NP hard in general. Fortunately, we can use a heuristic algorithm based on Max-Min Ant System (MMAS) to compute it approximately \cite{Zhang2018Structural}. The detailed MMAS scheme is provided in Section \ref{sec:3}.

\subsection{Graph Representation Learning}

Graph is a common and convenient way to represent  relationships among entities. The representation learning or feature learning studies how to convert the network information to dense, low-dimensional, real-valued vectors, which are further used as the input features in downstream machine learning tasks. This step has attracted great attention since techniques in representation learning allow a system to automatically discover the features from raw data, and transfer network information to quantitative relations.

In particular, the input of this step is a graph and the outputs are series of vectors associated to each node on the graph. The similarity between two nodes is determined by both the similarity of their neighborhood nodes and the weights along the path to those nodes. A good representation can characterize this similarity in the sense that the more similar two nodes are, the closer two associated vectors will be. We briefly introduce the method DeepWalk \cite{Perozzi2014Deepwalk}, which is one of the first widespread representation learning methods and is the method that we apply on the behavior graph.

DeepWalk uses local information obtained from several random walks with certain length on the graph to learn latent representations, and treats node sequences as the equivalent of sentences. It utilizes the SkipGram algorithm \cite{Mikolov2013Efficient} and Hierarchical Softmax model \cite{Morin2005Hierarchical, Mnih2009scalable} to probabilize the pairs of nodes in each local window in the random walk sequence, and finally maximizes the likelihood by stochastic gradient descent (SGD). SkipGram is a language model that maximizes the co-occurrence probability among the words.  It maximizes classification of a word based on another word in the same sentence. More precisely, it uses each current word as an input of a log-linear classifier with continuous projection layer, and predicts words within a certain range before and after the current word \cite{Mikolov2013Efficient}.

\subsection{Problem Formulation}\label{sec:formulation}

We now formulate our interactive news recommendation problem. For a given user $i$, we denote the behavior history set by $b_i = \{b_{i,1}, b_{i,2}, \ldots, b_{i, {\T_i}}\}$, where $b_{i,j}$ for $j = 1,\ldots, \T_i$ is the $j$-th behavior triggered by user $i$, and $\T_i$ is the total number of user's behaviors. Here ``trigger'' means user has browsed one piece of news so that one of five positive behaviors, from click to share, has been generated. Thus, each behavior $b_{i,j}$ corresponds to a piece of news, say $d_{i,j}$. For simplicity, we assume that for each user $i$ the triggered news $\{d_{i,j}\}_{j=1}^{\T_i}$ are different from each other. In practice, if a user generates two different behaviors on the same news, we will only keep the most positive behavior (another option is to keep the most recent behavior).

{\red
\begin{remark}
We order the behaviors from unclick, click, like, follow, comment, to share. We think the behaviors in this order are more and more positive. In Section \ref{sec:3.1}, we assign different weights to the edges between user and news nodes that are induced by different behaviors. The weights reflect the connection intensity of two nodes. The more positive the behavior, the larger the weight, and the stronger the connection between two nodes. Thus, if a user triggers different behaviors on the same news, we think the extent that the user is interested in this news is decided by his/her most positive behavior, which is a reasonable presumption.

We also mention that doing a reduction for repeated samples is necessary for our method, since we do not~deal with multiple edges between two nodes. In practice, it is also rare to have multiple behaviors on the same news (since news is updated instantly). So our data adjustment has a negligible effect on the recommendation. We emphasize that only if we have repeated (user, news) pair do we do an adjustment. Most of behaviors are preserved. Certainly, designing a method for dealing with repeated samples is an interesting future work.

\end{remark}
}

Moreover, in our scenario, each news $d$ is composed of a sequence of words indicating contents, tags, primary and secondary categories, i.e.
\begin{equation}\label{equ:news:vector}
d = [w_1,w_2, \ldots; tag_1, tag_2,\ldots; cat_1,cat_2].
\end{equation}
In Section \ref{sec:3}, we will discuss how to learn an embedding matrix of the above vector to further obtain the news representations via a convolutional network. We should mention that \eqref{equ:news:vector} does not include topics. We realize in our real dataset that topics are missing for some of news. Thus, we only use common features of news to construct their embedding matrices. However, as one type of nodes on the interaction behavior graph, topics will affect the learned embeddings implicitly. Also note that each news $d$ may be associated with different users on the interaction behavior graph.

Given the behavior history sets for all users, we aim to predict whether user $i$ will behave positively on a candidate news that he/she has not ever seen before. Further, what is the probability of each of five possible behaviors (from click to share).

\section{The Proposed Method}\label{sec:3}

In this section, we introduce our method called graph-based behavior-aware network (GBAN). We decompose GBAN into five steps introduced in the next five subsections respectively: interaction behavior graph construction, graph representation learning, user concentration feature learning, user and news representations, and behavior prediction.

\begin{figure*}[!htbp]
\centering
\includegraphics[scale=0.35]{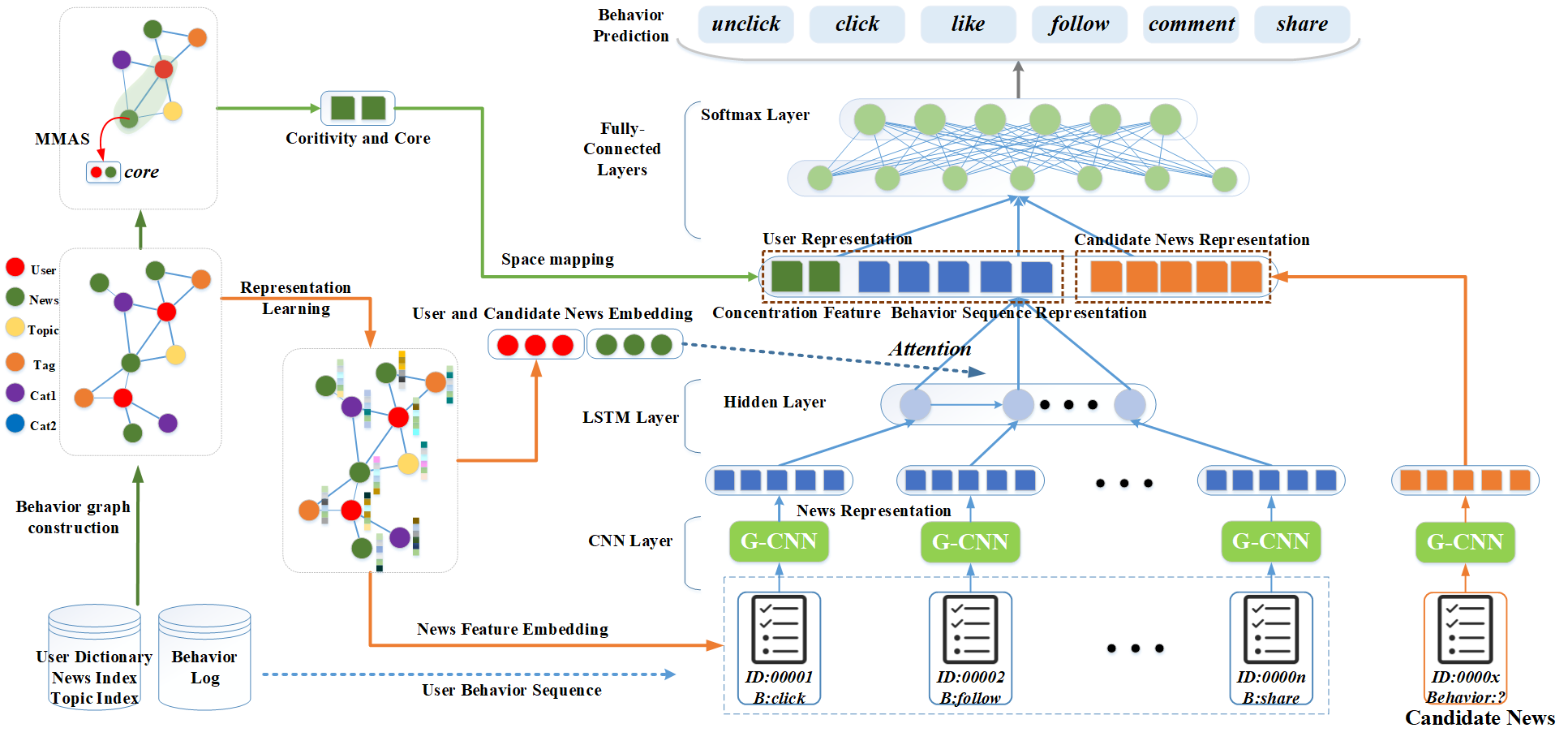}
\caption{The workflow of GBAN. From the bottom left part to the top right part, GBAN constructs an interaction behavior graph, calculates core and coritivity and performs graph representation learning in parallel, learns news representations via G-CNN, learns behavior sequence representations via LSTM, and feeds user representations and candidate news representations to two fully-connected layers to predict probabilities of six behaviors.}
\label{fig:workflow}
\end{figure*}

As illustrated in Figure \ref{fig:workflow}, our recommendation system predicts the probabilities of six behaviors of a user on a candidate news. First, we construct an interaction behavior graph based on users profiles and behaviors logs. Second, we obtain the nodes embeddings from graph representation learning, and meanwhile calculate core and coritivity for subgraphs induced by each user as his/her concentration feature. Third, we design a G-CNN to learn news representations, and an attention-based LSTM to learn the behavior sequence representations, which together with concentration feature constitute user representations. Finally, we combine user representations with candidate news representations, and feed them into a two-layer fully-connected neural network to perform the six-classification task.

\subsection{Construction of Interaction Behavior Graph}\label{sec:3.1}

To fully exploit the diversified behaviors and the associations between users and news, we construct a weighted heterogeneous graph for the interactive news data. We use Tencent Interactive News data to illustrate the construction. Other data with multiple behaviors share the same construction process. For example, we also study MovieLens data \cite{Harper2015movielens} in the experiment in Section \ref{sec:4}.

\vskip5pt

\noindent\textbf{Tencent data description:} This dataset comes from the service log of Tencent News App. It consists of four files: behavior log, user dictionary, news index, and topic index. Each line of the behavior log contains user ID, timestamp, news ID, topic ID, and user's behavior. Each line of the user dictionary contains user ID, a primary category ID, a secondary category ID and multiple tags. Moreover, for each user, categories and tags labels are assigned with weights (i.e. proportions) calculated according to his/her browsing history. The other two files have similar formats to the user dictionary. In news index file, each news ID corresponds to a topic ID, primary and secondary categories IDs, multiple tags and a sequence of content words; while in topic index file, each topic ID corresponds to primary and secondary categories IDs and multiple tags. Here, categories and tags also come with prespecified weights.

Given such a dataset, we construct the interaction behavior graph as follows. We consider five types of nodes: user, news, category, tag, and topic. We denote them by $U$, $D$, $C$, $Tg$, $T$, respectively. Then, the graph $G$ is given by $G = (V, E)$, where $V = U\cup D\cup C\cup Tg\cup T$ is the node set and $E = \{(v_{i,1}, v_{i,2}, \omega_i)\}_i$ is the edge set. The elements of $E$ are triple, with $v_{i,1}$, $v_{i,2}$ being two corresponding nodes and $w_i$ being the assigned weights.

The weights are assigned according to different edges types. In particular, the weight of \textbf{user-news} edge is assigned based on user's behavior. We let $\{\text{click}: 0.5, \text{ like}: 0.6, \text{ follow}: 0.7, \text{ commend}: 0.8, \text{ share}: 1\}$. If there is no edge between a user node and a news node, that means the user unclick the news. For \textbf{user-category/tag} edge, the weight is given by user dictionary file, which is essentially the proportion of each news category/tag in his/her browsing history. Similarly, we manually calculate the proportion of each news topic in the browsing history, and assign weights accordingly for \textbf{user-topic} edge. Moreover, the weight of \textbf{news-category/tag} edge is given by news index file and the weight of \textbf{topic-category/tag} edge is given by topic index file. Lastly, since each news belongs to a unique topic, we set weight to be $1$ for \textbf{news-topic} edge. A summary of weight assignment is displayed in Table \ref{tab:node}.

\begin{table}[!t]
	\caption{Summary of weights of the behavior graph.}
	\scalebox{0.95}{
		\setlength{\tabcolsep}{0.1mm}{
			\begin{tabular}{| c | c | c | c | c | c |}
				\hline
				Node & User & News & Category & Tag & Topic\\
				\hline
				\multirow{5}{*}{User} & \multirow{5}{*}{$\times$} & \multicolumn{1}{|l|}{click: 0.5} & \multicolumn{2}{c|}{\multirow{5}{*}{\shortstack{provided\\ in user\\ dictionary}}} & \multirow{5}{*}{\shortstack{proportion of \\ each topic in\\ behavior log}} \\
				& & \multicolumn{1}{|l|}{like: 0.6} & \multicolumn{2}{c|}{} & \\
				& & \multicolumn{1}{|l|}{follow: 0.7} & \multicolumn{2}{c|}{} & \\
				& & \multicolumn{1}{|l|}{comment: 0.8} & \multicolumn{2}{c|}{} &\\
				& & \multicolumn{1}{|l|}{share: 1} & \multicolumn{2}{c|}{}  & \\
				\hline
				\multirow{2}{*}{News} & \multirow{2}{*}{} & \multirow{2}{*}{$\times$} & \multicolumn{2}{c|}{\multirow{2}{*}{\shortstack{provided in\\ news index}}} & \multirow{2}{*}{1}\\
				& & & \multicolumn{2}{c|}{} & \\
				\hline
				Category & & & \multicolumn{2}{c|}{\multirow{2}{*}{$\times$}} & \multirow{2}{*}{\shortstack{provided in\\ topic index}}\\
				\cline{0-0}
				Tag && &\multicolumn{2}{c|}{}  &\\
				\hline
				\multicolumn{5}{l}{Note: $\times$ means there is no edge between nodes.}
		\end{tabular}}
	}
	\centering
	\label{tab:node}
\end{table}

\begin{remark}

The reason why we assign weights in this way is because that the weight between two nodes should reflect their association extent. This association extent can be further reflected from the behaviors or the browsing history. Intuitively, behaviors from unclick to share is becoming more and more positive, which implies that this user is more and more associated with this piece of news. Similarly, the association between user and topic is determined by the proportion of news of this topic in the browsing history. Thus, we set the weight accordingly. When we perform graph representation learning, the transition probability between nodes is proportional to the weight. Then, two nodes are more likely to appear in the same random walk if they have higher weight, and the learned embeddings are also more similar.

\end{remark}

A detailed behavior graph construction process is presented in Algorithm~\ref{alg:BG}. The complexity of Algorithm~\ref{alg:BG} equals to the total number of lines in four data files. Different from the existing heterogeneous graphs \cite{2019HRec,2013Link}, our constructed behavior graph contains rich nodes information such as topic, tag, and category, aside from user and news. This graph structure is more suitable for interactive recommendation, and allows us to increase the breadth of features. For example, we are able to study the subgraph induced by each user and make recommendation personalized by involving the concentration feature. The different behaviors are considered when assigning weights, which provides a way to extend the application of heterogeneous graphs to interactive news scenario.

\begin{algorithm}[t]
\caption{Construction of Behavior Graph}	
\label{alg:BG}
\begin{algorithmic}[1]
\State \textbf{Input:} Behavior log, User dictionary, News index, Topic index;
\State Let $NodeSet, EdgeSet, TopicEdgeDict = \{\},\{\},\{\}$;
\State Let $BeDict = \{\text{unclick}: 0, \text{ click}: 0.5, \text{ like}: 0.6, \text{ follow}: 0.7, \text{ commend}: 0.8, \text{ share}: 1\}$;
\item[]
\For{each line $i$ in Behavior log}
\State Extract $user_i, news_i, topic_i, behavior_i$;
\State Add $user_i, news_i, topic_i$ to $NodeSet$;
\State Let $weight_i = BeDict\{behavior_i\}$;
\State Add triple $(user_i, news_i, weight_i)$ to $EdgeSet$;
\If{$user_i, topic_i$ not in $TopicEdgeDict$}
\State Let $TopicEdgeDict_{user_i,topic_i}=0$;
\EndIf
\State Let $TopicEdgeDict_{user_i,topic_i} += weight_i$;
\EndFor

\For {$user_i,topic_j$ in $TopicEdgeDict$}
\State Let $weight_{ij} = \frac{TopicEdgeDict_{user_i, topic_j}}{\sum_{topic} TopicEdgeDict_{user_i, topic}}$;
\State Add triple $(user_i, topic_j, weight_{ij})$ to $EdgeSet$;
\EndFor

\item[]
\For {$user_i$ in User dictionary}
\State Extract primary category $cat_{i,1}$, secondary category $cat_{i,2}$, and multiple tags $[tag_{i,1},\ldots,tag_{i,u_i}]$ with their weights $w_{cat_{i,1}}, w_{cat_{i,2}}, w_{tag_{i,1}},\ldots,w_{tag_{i,u_i}}$;
\State Add $user_i$, $cat_{i,1},cat_{i,2},\{tag_{i,k}\}_{k=1}^{u_i}$ to $NodeSet$;
\State Add triples $(user_i, cat_{i,1}, w_{cat_{i,1}})$,$(user_i, cat_{i,2}, w_{cat_{i,2}})$, $\{(user_i, tag_{i,k}, w_{tag_{i,k}})\}_{k=1}^{u_i}$ to $EdgeSet$;
\EndFor

\item[]
\For {$news_i$ in News index}
\State Extract $topic_i$, categories $cat_{i,1}$ and $cat_{i,2}$, and multiple tags $[tag_{i,1},\ldots,tag_{i,n_i}]$ with their weights $1,w_{cat_{i,1}},w_{cat_{i,2}}, w_{tag_{i,1}},\ldots,w_{tag_{i, n_i}}$;
\State Add $news_i, topic_i, cat_{i,1}, cat_{i,2}, \{tag_{i,k}\}_{k=1}^{n_i}$ to $NodeSet$;
\State Add triples $(news_i,topic_i,1)$, $(news_i, cat_{i,1}, w_{cat_{i,1}})$, $(news_i,cat_{i,2},w_{cat_{i,2}})$, $\{(news_i, tag_{i,k}, w_{tag_{i,k}})\}_{k=1}^{n_i}$ to $EdgeSet$;
\EndFor

\item[]
\For {$topic_i$ in Topic index}
\State Extract categories $cat_{i,1}$ and $cat_{i,2}$, and multiple~tags $[tag_{i,1}, \ldots, tag_{i,t_i}]$ with their weights $w_{cat_{i,1}}, w_{cat_{i,2}}$, $w_{tag_{i,1}}, \ldots, w_{tag_{i,t_i}}$;
\State Add $topic_i, cat_{i,1}, cat_{i,2}, \{tag_{i,k}\}_{k=1}^{t_i}$ to $NodeSet$;
\State Add triples $(topic_i, cat_{i,1}, w_{cat_{i,1}}), (topic_i,cat_{i,2},w_{cat_{i,2}})$, $\{(topic_i,tag_{i,k},w_{tag_{i,k}})\}_{k=1}^{t_i}$ to $EdgeSet$;
\EndFor

\State \textbf{Output:} $NodeSet, EdgeSet$.
\end{algorithmic}
\end{algorithm}

\subsection{Graph Representation Learning}\label{subsec:grl}

Given the constructed behavior graph, we then perform graph representation learning to embed each graph node a vector. In this step we ignore the heterogeneity of the graph (i.e. treat all types of nodes equally), and apply DeepWalk algorithm \cite{Perozzi2014Deepwalk} on the graph.

In fact, the standard DeepWalk has to be slightly adjusted since it is designed for unweighted graphs. In our case, when we generate random walks, we let the transition probability be proportional to the weight. Thus, the higher the weight, the more likely two nodes will appear in the same node sequence. Noting that the weight is increased when the behavior varies from click to share, DeepWalk ensures that the shared news will be learned as the strongest evidence of the preferences. Such tendency is consistent with the users habits. It is also useful to note that all learned embeddings naturally integrate the user's behaviors information and news features (e.g. topics, tags) information.

We briefly demonstrate the implementation of the adjusted DeepWalk. We have two steps. First, for each node $v_i$, we generate multiple random walks starting from $v_i$ with length $t$. We donate a random walk starting from $v_i$ as $\{v_i^0=v_i,v_i^1,\cdots,v_i^t\}$, where $v_i^j\; (j\geq 1)$ is a node chosen at random from the neighbors $\N(v_i^{j-1})$ of node $v_i^{j-1}$. Different from standard DeepWalk, we let transition probability from $v_i^{j-1}$ to $v_i^j$, denoted by $\{p_{ij}^v\}_{v\in \N(v_i^{j-1})}$, be $p_{ij}^v=\frac{weight_{v_i^{j-1}, v}}{\sum_{x\in \N(v_i^{j-1})}weight_{v_i^{j-1}, x}}$, and $weight_{v_i^{j-1}, v}$ is the weight of edge between $v_i^{j-1}$ and $v$. Second, following standard DeepWalk, we apply SkipGram algorithm \cite{Mikolov2013Efficient} to generate node representations for each random walk. We refer to Section~4 of \cite{Perozzi2014Deepwalk} for details on how to apply SkipGram on random walks.

\begin{remark}

We prefer DeepWalk for representation learning due to three reasons.
\begin{enumerate}[label=(\alph*),topsep=0pt]
\setlength\itemsep{0.0em}
\item Designing efficient methods targeting general heterogeneous graphs is still an open problem. Existing methods \cite{Ma2021AEGCN, Kipf2017Semi} cannot fit in our scenario and are computationally expansive. The number and the types of nodes and edges of our behavior graph change rapidly over time, which may lead to a delay for recommending news when using very complex methods. However, DeepWalk has clear advantages on parallelism and adaptability, and has been extensively validated in practice.

\item Ignoring heterogeneity of graph in representation learning step does not result in a worse performance. Conversely, DeepWalk is enough to let GBAN achieve state-of-the-art performance, and can be performed in real time.

\item Although other methods for homogeneous graph representation learning allow flexible embeddings, they bring additional issues. For example, node2vec \cite{Grover2016node2vec} involves two parameters in the transition probability: one prioritizes a breadth-first-search procedure and one prioritizes a depth-first-search procedure. These auxiliary parameters bring flexibility to the method, but also let the algorithm tend to generate biased random walks. The bias may affect the meaning of our weights. On the other hand, as the prototype of many other advanced methods, DeepWalk simply generates unbiased random walks based on our prespecified weights. It is simple, efficient, and can perfectly reflect the meaning of the weights.
\end{enumerate}	
To justify the above reasons, we compare our adjusted DeepWalk with the standard DeepWalk and node2vec methods in experiments (see Table \ref{tab:vgban}). We show our method enjoys a better performance.

\end{remark}

\begin{figure*}[!t]
	\centering
	\includegraphics[scale=0.15]{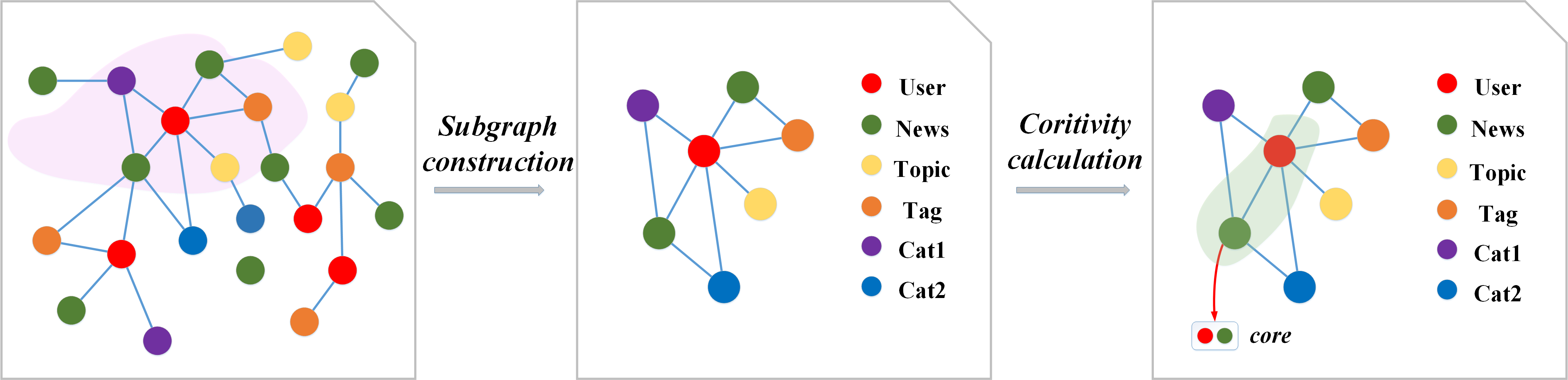}
	\caption{The concentration feature learning.}
	\label{fig:3}
\end{figure*}

\subsection{Concentration Feature Learning}\label{subsec:ucfl}

Users' concentration degree directly affects the pattern of their behaviors. Intuitively, users tend to repeat similar behaviors for a certain class of news if they have relatively narrow preferences. In contrast, they will be more likely to behave in a different way if their preferences are extremely wide. In order to personalize the recommendation, we take user's concentration feature into account.

As introduced in Section \ref{sec:2}, core and coritivity are used to describe the importance of a group of nodes on the graph and are widely used in information security and brain graph structure analysis \cite{Zhang2018Structural}. We also use them to characterize the user's concentration degree.

As illustrated in Figure \ref{fig:3}, we first extract subgraphs from the full behavior graph induced by each user. Each subgraph contains only the specific user node and its all adjacent nodes. We calculate core $S$ and coritivity $h(G)$ for the subgraph, and the pair $(|S|, h(G))$ will be used to measure the concentration degree of this user. Since $S$ is not unique, we will choose $S$ such that $|S|$ is as small as possible. The reasonability of this measure comes from the definition of core and coritivity.

Since computing core and coritivity is NP-hard in general, we apply Max-Min Ant System (MMAS) algorithm to approximately compute them. MMAS \cite{Stuetzle1998Improvements, Stuetzle2000MAX, Zhang2018Structural} is a random search algorithm that simulates the ants foraging behavior in nature. When ants are foraging, they secrete chemical hormone, called pheromone, on the path they pass by, and meanwhile judge their direction according to the concentration of nearby pheromone. The key component of MMAS is a parameterized probability model called a pheromone model. In each iteration, the system generates a population based on the pheromone model, and then updates the pheromone model by the generated population information to affect the next-generation population. The details are presented in Algorithm \ref{alg:1}. The total number of iterations of MMAS is $maxDepth \times Epoch$. In practice, letting the product be $O(10^2)$ is good enough. We then map coritivity and core to the Euclidean space using
\begin{equation}\label{equ:map1}
cf = \texttt{ReLu}(W_{map_1} [ Coritivity,Core] + \beta_{map_1})\in\mR^2
\end{equation}
where $W_{map_1}\in \mR^{2\times 2},\beta_{map_1}\in \mR^2$. Finally, we treat $cf$ as the concentration feature representation of the user.

\begin{algorithm}[t]
\caption{Max-Min Ant System (MMAS)}
\label{alg:1}
\begin{algorithmic}[1]
\State \textbf{Input:} Subgraph $SubG = (SubV, SubE)$, maximum iteration depth $maxDepth>0$, the number of epochs $Epoch>0$, the pheromone volatilization coefficient $\rho \in (0,1)$, and scalars $\tau_{min}, \tau_{max}, \alpha, \beta>0$;
\State Let $Coritivity = 0, Core = \{\}$;
\State Let $\tau=\{\tau_i\}_{i \in SubV}, \eta=\{\eta_i\}_{i\in SubV}$ such that $\tau_i= \tau_{min}, \eta_i = 1/d_i$ where $d_i$ is the degree of node $i$ in $SubV$;
\For{$k=0,1,\ldots, Epoch$}
\State $temCoritivity = 0$, $temCores = \{\}$;
\State Calculate node selection probabilities $P=\{P_i=\frac{\tau_i^{\alpha}\eta_i^{\beta}}{\sum_{j\in SubV}\tau_j^{\alpha}\eta_j^{\beta}}: i \in SubV\}$;

\For{$depth = 0,1,\ldots, maxDepth$}
\State Generate $numCore$ uniformly from $1$ to $|SubV|$, where $numCore$ is the number of nodes in the core set;
\State Allocate the core set $S$ according to probabilities $P$ such that $|S|=numCore$;
\State Compute $newCoritivity = \omega(G-S)-|S|$;

\If {$newCoritivity > temCoritivity$}
\State $temCoritivity = newCoritivity$;
\State $temCore =  S$;
\EndIf
\EndFor

\If{$temCoritivity > Coritivity$}
\State $Coritivity = temCoritivity$;
\State $Core = temCore$;
\EndIf

\State Compute $\Delta \tau^{best}= \frac{temCoritivity}{Coritivity}$;
\For{node $i$ in $temCore$}
\State $\tau_i = [(1-\rho)\tau_i+\Delta \tau^{best}]_{\tau_{min}}^{\tau_{max}}$ ($[x]_{a}^b$ means $x=a$ if $x<a$; $x=b$ if $x>b$; $x=x$ if $a\leq x \leq b$);
\EndFor
\EndFor
\State \textbf{Output:} $Coritivity, Core$.
\end{algorithmic}
\end{algorithm}

\subsection{News and Users Representation Learning} \label{subsec:nur}

After node embedding in Section \ref{subsec:grl} and concentration feature learning in Section \ref{subsec:ucfl}, we now set the stage to learn news and users (low-dimensional) representations, which are used as inputs of the downstream classification task. We divide the whole process to four steps: assemble news features, design a CNN to learn news representations, design an attention-based LSTM to learn behavior sequence representations, and assemble users representations.

\vskip 5pt
\noindent \textbf{Step 1: assemble news features.} As introduced in \eqref{equ:news:vector}, each news $d$ is composed of a sequence of content words, multiple tags, and primary and secondary categories. Correspondingly, the news features are composed by the embeddings of these components.

Tags and categories are nodes of the behavior graph, and their embeddings are obtained from graph representation learning in Section \ref{subsec:grl}. For simplicity, for each news we only use embeddings of $s$ tags that have the largest weights. To learn embeddings for content words, we perform Chinese word segmentation\footnote{It is not needed for English news.} and extract $M$ keywords by Jieba Chinese text segmentation\footnote{\url{http://pypi.python.org/pypi/jieba/}.} and TF-IDF keyword extraction \cite{Juanzi2007Keyword}. Then, the embedding of each keyword is from the word2vec library\footnote{\hskip-1pt\url{https://github.com/Embedding/Chinese-Word-Vectors}.}. However, the word embedding and graph embedding are learned in different spaces, so we map the word embedding from the word space to the Euclidean space using
\begin{equation}\label{equ:1}
tran\_w_{i} = \texttt{tanh}(W_{map_2} w_{i} + \beta_{map_2}),\quad\; i = 1, \ldots, M,
\end{equation}
where we abuse the notation of weights to let $w_i$ denote the keyword embedding vector, and $W_{map_2}\in \mR^{n\times n_w}$ and $\beta_{map_2}\in\mR^n$ are trainable transformation matrix and bias with compatible dimensions. Finally, the keywords embeddings together with nodes embeddings of tags and categories constitute the news feature matrix:
\begin{align}\label{equ:2}
news_{in}(d)=&[tran\_w_{1}, \ldots, tran\_w_{M}, E\_tag_{1}, \ldots, E\_tag_{s}, \nonumber
\\ &E\_cat_{1}, E\_cat_{2}] \in \mathbb{R}^{n\times (M+s+2)}.
\end{align}
Here, $E\_tag_i$ and $E\_cat_i$ are tag and category embeddings from graph representation learning in Section \ref{subsec:grl}.

\vskip 5pt
\noindent \textbf{Step 2: news representations learning.} We then feed the feature matrix \eqref{equ:2} into a CNN to learn news representations. We use multiple filters $C\in  \mathbb{R}^{n \times l \times k_1}$ with varying window sizes $l$ and, inspired by DKN~\cite{Wang2018DKN}, a max-pooling operation on the tensor. We thus have
\begin{align}\label{m_news}
m_d = \texttt{maxpooling}(f(news_{in}(d)\otimes C + \beta\boldsymbol{1}^T )) \in \mR^{k_1},
\end{align}
where $\otimes$ is the convolution operator, $\beta\in \mR^{M+s+2}$ is bias vector, $\boldsymbol{1}\in\mathbb{R}^{k_1}$ is all-one vector, and $f$ is $\texttt{ReLu}$ and is evaluated entrywisely. We emphasize that we use the same CNN layer for all news, so that the filters $C$ do not depend on the input news.

\vskip 5pt
\noindent \textbf{Step 3: behavior sequence representations learning.} Based on the learned news representations, we design an attention-based LSTM to further learn behavior sequence representations. The input of LSTM is a sequence of representations of news that is triggered by the user. Recall from Section \ref{sec:formulation} that,  for a specific user, the behavior sequence is $\{b_{1}, b_{2}, \ldots, b_{\T}\}$. Each behavior $b_j$ correspond to a piece of news $d_j$. In experiment, we split the long news sequence $\{d_{1}, d_{2}, \cdots, d_{\T}\}$ into short sequences so that we have multiple samples even for one user. For example, we can get $\T-r$ samples $\{(d_q,\ldots, d_{q+r-1}), d_{q+r}, b_{q+r}\}_{q = 1}^{\T-r}$. For each sample, we use the previous $r$ news to predict the behavior on the $(r+1)$-th news. We describe the behavior sequence representations learning on the first sample $(q=1)$, and $[m_{d_1},m_{d_2},\cdots,m_{d_r}]$ is the news representation matrix.

At step $j$ of LSTM, we use $m_{d_j}$ to update the current cell state $c_j\in\mR^{k_2}$ and hidden state $h_j\in\mR^{k_2}$ by
\begin{align}\label{LSTM}
i_j &= \texttt{Sigmoid}(W_1 m_{d_j}+W_5 h_{j-1}+\beta_1),\\
f_j &= \texttt{Sigmoid}(W_2 m_{d_j}+W_6 h_{j-1}+\beta_2),\\
o_j &= \texttt{Sigmoid}(W_3 m_{d_j}+W_7 h_{j-1}+\beta_3),\\
\hat{c}_j &= \texttt{tanh}(W_4 m_{d_j}+W_8 h_{j-1}+\beta_4),\\
c_j&=f_j\odot c_{j-1}+i_j\odot \hat{c}_j,\\
h_j&= o_j\odot \texttt{tanh}(c_j).
\end{align}
where $i,f,o$ are gate activations, $\odot$ meas elementwise multiplication, $\{W_i\}_{i=1}^4\in \mR^{k_2\times k_1}$, $\{W_i\}_{i=5}^{8}\in \mR^{k_2\times k_2}$, $\{\beta_i\}_{i=1}^4 \in \mR^{k_2} $ are trainable weight matrices and biases.

Then, we employ the following attention mechanism to calculate the weighted behavior representation $s$:
\begin{equation}\label{equ:ann}
s =  \frac{\sum_{i=1}^r \exp{(e(h_i, u, p))} \cdot h_i}{\sum_{j=1}^r\exp{(e(h_j,u,p))}} \in \mR^{k_2}
\end{equation}
with
\begin{align}
e(h_i, u, p)=& \beta_b^T\texttt{tanh}(W_Hh_i+W_Uu+W_Pp+\beta_a).
\end{align}
Here, $e$ is a score function, which scores the importance of news for composing the behavior representation; $u\in \mR^n$ is the user node embedding, and $p\in\mR^n$ is the candidate news node embedding (here $p = E\_d_{r+1}$ as $d_{r+1}$ is our candidate news, see notations in \eqref{equ:2}). Both of them come from graph representation learning in Section \ref{subsec:grl}. $W_U, W_P \in \mR^{n \times n}$ and $W_H \in \mR^{n \times k_2}$ are weight matrices; and $\beta_a,\beta_b\in \mR^{n}$ are bias vectors.

\begin{remark}\label{rem:1}

Our model is simpler than DAN \cite{Zhu2019DAN} which also adopted an attention-based CNN-LSTM framework.
\begin{enumerate}[label=(\alph*),topsep=0pt]
\setlength\itemsep{0.0em}
\item In DAN, the authors designed an attention-based LSTM called ARNN. ARNN involves attention mechanism on each state of LSTM. In particular, after we feed sequence $[m_{d_1},m_{d_2},\cdots,m_{d_r}]$ into LSTM, we get states $[h_1,h_2,\cdots,h_r]$. For each $i = 2, 3,\ldots, r$, ARNN involves an attention network ANN for $h_i$ to provide attention weights for all past LSTM states $[h_1, \ldots, h_{i-1}]$. In total, ARNN has $r-1$ ANN layers to obtain $r-1$ sequential features, and integrates them as behavior sequence representations. In our model, we only perform one attention network as shown in \eqref{equ:ann}.

\item In DAN (and DKN \cite{Wang2018DKN}), the attention network is applied on the embeddings of the clicked news in the browsing history, while our attention network is applied on the embeddings of the user node and candidate news node. This difference allows us to learn personalized, and candidate news-oriented user representations.

\end{enumerate}	

\end{remark}
\vskip 5pt

\noindent \textbf{Step 4: assemble user representations.} We are now able to construct user representations, which consist of the concentration feature $cf$ from \eqref{equ:map1} and the behavior sequence representations $s$ from \eqref{equ:ann}. Finally, combining the user representations with the candidate news representations from Step 2, we obtain the input of the downstream classification task.

\subsection{Behavior Prediction}

By far, we have illustrated how to obtain user and news representations. We now consider the behavior prediction: how the user will behave on an unbrowsed, candidate news. We formalize the ranking problem as a six-classification problem. We believe the six-classification task is more appropriate for interactive news recommendation than CTR prediction.

We design a two-layer fully-connected network to conduct classification task. The input is user and candidate news representations $[cf,s,m_{d_{r+1}}]$. Then we apply
\begin{align}\label{equ:fc}
h_{fc} &= \texttt{ReLu}(W_{fc}[cf,s,m_{d_{r+1}}]+\beta_{fc}),\\
\theta &= W_{output}h_{fc}+\beta_{output} \in \mR^{|B|}.
\end{align}
where $W_{fc}\in \mR^{k_3\times (k1+k2+2)}, W_{output} \in \mR^{|B|\times k_3}$, $\beta_{fc}\in \mR^{k_3},\beta_{output} \in \mR^{|B|} $ are trainable weight matrices and biases. The loss function is
\begin{equation}\label{loss:1}
Loss=-\frac{1}{N}\sum^N_{k=1}\sum^{|B|}_{i=1}y^k_i\times \log(\hat{y}_i^k),
\end{equation}
with
\begin{equation}\label{equ:prob}
\hat{y}_i^k= \texttt{softmax}(\theta^k)=\exp(\theta_i^k)/\sum_j{\exp(\theta_j^k)}.
\end{equation}
Here, $N$ is the number of samples in the training set, $|B|$ is the number of behavior types (in our scenario $|B| = 6$), $\theta^k$ is the output of last layer for the $k$-th samples, $y^k = (y_1^k,\ldots,y_{|B|}^k)$ is the true probability distribution, i.e. if the user triggers the $l$-th behavior, then $y^k_l = 1$ and $y^k_j = 0, \forall j\neq l$.

As standard in the literature, we run mini-bath Adam to train all parameters. The setup of Adam including learning rate and batch size are introduced in Section~\ref{sec:4}.

\section{Experiment}\label{sec:4}

In this section, we test the effectiveness of our proposed recommendation system GBAN by offline evaluation. We first introduce two datasets in Section \ref{sec:4.1}, then introduce model evaluation metrics in Section \ref{sec:4.2}. The experimental results on two datasets, with comprehensive comparisons with other state-of-the-art models, are presented in Sections \ref{sec:4.3} and \ref{sec:4.4}, respectively. We also show that all our techniques including adjustment of DeepWalk, concentration feature, and attention mechanism are useful to enhance the performance. Our code is publicly available at \url{https://github.com/AI-luyuan/GBANRecommendation}. The implementation is performed on a computer with 4.2GHz Intell Core i7 CPU, 32 RAM and GeForce GTX 1080Ti GPU.

\subsection{Dataset Description}\label{sec:4.1}

Our paper considers two datasets. One is Tencent Interactive News data and the other is MovieLens data \cite{Harper2015movielens}.
{\red
Both data are time series; hence we simply pick a time stamp to split the data into a training set and a test set: the data before the stamp consist of the training set and the data after the stamp consist of the test set. No resampling strategies are used in the experiment. We train parameters on the training set and then do evaluation (based on trained parameters) on the test set. In later presentation, for the same fixed training set, we generate 5 random initializations for weight matrices when we perform Adam algorithm, which is the same as \cite{Zhu2019DAN, Wang2018DKN}.
}

%

\begin{figure*}[tp]
\centering
\includegraphics[width=7in]{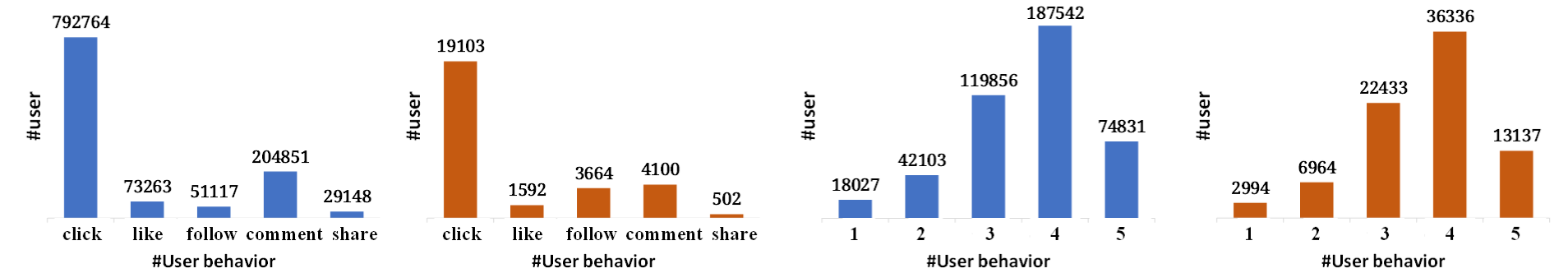}
\caption{Distribution of users' behaviors of Tencent data and MovieLens data. The left two figures correspond to Tencent data, and the right two figures correspond to MovieLens data. The blue figures correspond to the training set, and the yellow figures correspond to the test set.}
\label{fig:6} 
\end{figure*}

%
%


\vskip 5pt
\noindent\textbf{Tencent data:} Tencent data come from the past server logs of Tencent News App. It's a real industrial data but has not been publicly available. We have described the data, which include four log files, in Section \ref{sec:3.1}. 
In our experiment, we approximately sample 220k users in total, and collect data from June 19, 2018 to July 16, 2018 as training set and July 17, 2018 as test set. There are 4,604,572 logs in the training set and 289,610 logs in the test set. The basic statistics of the dataset are summarized in Table \ref{tab:1}, and the number of each type of positive behavior is shown in Figure \ref{fig:6}.

\begin{table}[tp]
\renewcommand\arraystretch{1.5}
\centering
\caption{ Tencent news data statistics}
\label{tab:1}
\setlength{\tabcolsep}{2.7mm}{
	\begin{tabular}{l|r|l|r}
		\hline
		\#users  &223,931 & \#tags & 29,564 \\
		\hline
		\#news & 54,338 & \#topics & 979   \\
		\hline
		\#logs & 4,894,182 & \#categories &  333 \\
		\hline
		\#nodes &300,932 &  \#edges & 10,072,780 \\		
		\hline
\end{tabular}}
\end{table}

From Table \ref{tab:1}, we see that there are about 10 million edges on the behavior graph connecting all kinds of nodes. From Figure \ref{fig:6}, we know that each behavior has a non-negligible proportion. Thus, it is important to consider all types of behaviors.

\vskip5pt
\noindent\textbf{MovieLens data:} To show the broad applicability of GBAN, we test it on a publicly available dataset called MovieLens data, in specific, hetrec2011 movielens-2k-v2 data\footnote{\url{https://grouplens.org/datasets/movielens/.}}. In such a dataset, each movie is associated with a title, a genre, a directory, a country, and multiple actors. We only extract actors with ranking 1. Each user rates a sequence of movies in a time series. To fit in our model, we convert ratings of $\{0.5, 1\},\{1.5, 2\}, \ldots, \{4.5,5\}$ into five types of {\red positive} behaviors, say $\{1,2,3,4,5\}$. We set a time $\T$, and select users with more than nine behaviors before $\T$ to form the training set. The selected users with their behaviors after $\T$ forms the test set.

Similar to Tencent data, we construct a behavior graph to summarize the data. We have six types of nodes: user, movie, genre, director, country, and actor; and have five types of edges: (user, movie), (movie, genre), (movie, director), (movie,country), (movie,actor). The (user, movie) edge weight is assigned according to the behavior: $i$ has weight $i/5$. The other edge weights are 1. The basic statistics of MovieLens data are summarized in Table \ref{tab:m}, and behavior distribution is shown in Figure \ref{fig:6}.

{\red
\begin{remark}

The similarity between Tencent data and MovieLens data is that the users' interests have different levels, revealed either by behaviors or rating scores. It's reasonable to say that the more positive behavior or the higher rating score the user has, the more interested the user is in the news/movie. The proposed method is not limited to having behaviors with real meaning. As long as the users' interests have different levels, we can always construct the corresponding graph (the extent differences are reflected by weights), and our method is still applied. However, it is worth mentioning that movie recommendation is sometimes treated as a different recommendation scenario due to the specific rating labels.

\end{remark}
}

\begin{table}[tp]
    \renewcommand\arraystretch{1.5}
	\centering
	\caption{MovieLens data statistic}
	\label{tab:m}
	\setlength{\tabcolsep}{2.7mm}{
	\begin{tabular}{l|r|l|r}
		\hline
		\#users  &848 & \#genres & 19 \\
		\hline
		\#movies & 10,197 & \#directors & 4,061   \\
		\hline
		\#countries & 71 & \#actors &  4,380 \\
		\hline
		\#nodes &19,576 & \#edges & 490,767  \\		
		\hline
	\end{tabular}}
\end{table}

\subsection{Model Evaluation}\label{sec:4.2}

As mentioned in Section \ref{subsec:nur} Step 3, each user has a behavior sequence $\{b_1, \ldots, b_\T\}$ with a news/movies sequence $\{d_1, \ldots, d_\T\}$. For each sequence, we generate~$\T-r$ behavior subsequences $\{(b_q, \ldots, b_{q+r-1})\}_{q = 1}^{\T-r}$ and corresponding news/movies subsequences $\{(d_q, \ldots, d_{q+r-1})\}_{q = 1}^{\T-r}$. For subsequence $q$, we use it to predict the user's behavior on the $(q+r)$-th news/movie. Thus, our (positive) samples are formed by $\{(d_q, \ldots, d_{q+r-1}), d_{q+r}, b_{q+r}\}_{q=1}^{\T-r}$. However, since behavior sequences only contain positive behaviors (click to share in news data and behaviors 1 to 5 in movie data), we manually generate negative samples as in \cite{Zhu2019DAN}. In particular, for each above sequence, we replace the positive candidate news/movie by randomly selecting a news/movie from the news/movies~set as the negative candidate.

To the best of our knowledge, there's no paper considering multiple-behaviors classification task for interactive recommendation scenario. To compare with other advanced models (which consider biclassification task, click v.s. unclick), we treat ``unclick'' behavior as negative class and the other behaviors as positive class. This is reasonable since, in our scenario, the other behaviors are also based on click behavior and, meanwhile, enhancing click-through rate (CTR) is also our ultimate goal. On the other hand, to fully exploit the benefits of our multi-classification task, we also use Cohen's kappa coefficient as one of metrics. In particular, we apply the following metrics to evaluate the performance of each competing model.

\vskip 5pt
\noindent\textbf{AUC metric:} This metric measures the area under the receiver operating characteristic (ROC) curve. It describes the probability that the model will rank positive samples in front of negative samples. AUC is not sensitive to the ratio of positive and negative samples. Thus, even in the case of sample imbalance, AUC is still a reasonable evaluation metric. In general, the larger AUC implies the better classification ability, and further implies the better performance.

\vskip5pt
\noindent\textbf{CTR metric:} As commonly used in news recommendation, our experiment on Tencent data also uses the precision as the offline surrogate of CTR, defined as
\begin{equation}
\text{precision} = \frac{\text{true positive}}{\text{true positive} + \text{false positive}}.
\end{equation}

\vskip 5pt
\noindent\textbf{Average precision (AP) metric:} For MovieLens data, we convert the original rating data format into an interactive news data format. However, as it's not a real interactive data, using precision as CTR metric, which tries to reflect the proportion of clicks in all recommended items, may not be suitable. We instead prefer to use AP, which calculates the area under the precision-recall curve without requiring a specific decision threshold. It reflects the trade-off between the accuracy of the classifier's recognition of positive samples and the ability to cover positive samples.

\vskip5pt
\noindent\textbf{Kappa coefficient metric:} The kappa coefficient $\kappa$ is an index used for consistency test, and can also be used to measure the effectiveness of classification. For classification problems, the consistency is whether the model prediction results are consistent with the actual classification labels. The formula is given by
\begin{equation}
\kappa = \frac{p_o-p_e}{1-p_e}
\end{equation}
with
\begin{align*}
p_0 &= \frac{a+d}{a+b+c+d},\\
p_c &= \frac{(a+c)\times (a+b)+(b+d)\times(c+d)}{(a+b+c+d)^2}.
\end{align*}
Here, $a$ is true positive, $b$ is false negative, $c$ is false positive, and $d$ is true negative.

To use the above four metrics, we first train GBAN model in Section \ref{sec:3} on the training set and derive optimal parameters. Then we apply optimal parameters on the test set to get probability distribution of behaviors $\hat{y}$ as in \eqref{equ:prob} for each candidate news. We thus have the predicted behavior that has the largest probability. Given the predicted labels and true labels, we can compute the above four metrics.

\subsection{Experiment on Tencent News Dataset}\label{sec:4.3}

\noindent{\bf Baseline:}
We implement five advanced methods listed in Table \ref{tab:2} as baselines in our experiments. Factorization machine (FM) \cite{Rendle2010Factorization} is a generic approach that allows to mimic most of factorization models through the feature engineering. There are two general FM-based models for recommendation: LibFM and DeepFM. LibFM \cite{Rendle2012Factorization} is a feature-based factorization model and widely used in CTR prediction. In our work, the input news features of LibFM have three parts: TF-IDF features \cite{Rajaraman2011Mining}, tags, and categories. We concatenate user and candidate news features as the input of LibFM. DeepFM \cite{Guo2017DeepFM} is a deep-learning-based model, which combines the power of factorization machines for recommendation and deep learning for feature learning in a novel neural network architecture. The input of DeepFM is the same as LibFM. In addition, YouTubeNet \cite{Covington2016Deep} is another architecture proposed recently. It is split into two parts according to the classic two-stage information retrieval dichotomy: a deep candidate generation model and a separate deep ranking model. Only the second part is applicable in interactive news recommendation problem. The last two recent algorithms are more related to our method. Okura et al. \cite{Okura2017Embedding} proposed an embedding-based method, which learns article representations based on a variant of de-noisy autoencoder, and then generates user representations by a RNN using browsing history as the input sequence. We denote this method by LSTM in Table \ref{tab:2}. In contrast, DKN \cite{Wang2018DKN} is a content-based deep recommendation model. We note that only the attention-based CNN part in DKN is suitable for our scenario.

\begin{table}[tp]
\renewcommand\arraystretch{1.5}
\centering
\caption{Baseline models descriptions}
\label{tab:2}
\setlength{\tabcolsep}{1.8mm}{
\begin{tabular}{l|l}
\hline
Model  &Description \\
\hline
LibFM \cite{Rendle2010Factorization} & Feature-based factorization mode  \\
\hline
DeepFM \cite{Rendle2012Factorization} & FM-based neural network \\
\hline
YouTubeNet \cite{Covington2016Deep} & Deep ranking network \\		
\hline
LSTM \cite{Okura2017Embedding} & Recurrent model using LSTM \\		
\hline
DKN \cite{Wang2018DKN} & Deep model using attention-based CNN\\		
\hline
\end{tabular}}
\end{table}

\vspace{5pt}
\noindent{\bf Parameter setup:} In our model, we let $r=5$, i.e. we use every $5$ triggered news to predict the next one for each user. For MMAS in Algorithm \ref{alg:1}, we let $Epoch = 50$, $maxDepth = 10$, $\rho=0.5$, $\tau_{min}=0.001$, $\tau_{max}=10$, $\alpha=0.2$ and $\beta=0.8$. For graph representation learning, we set the length of each random walk to be 20, and let $n = 300$ (i.e. the dimension of each node embedding is 300, see Section \ref{subsec:nur} \eqref{equ:2}). For the dimensions of hidden layers in LSTM and CNN, we set $l=3$, $k_1= k_2=50$, and $k_3=10$. We apply mini-batch Adam for adjusting the learning rate. The batch size is 1200 and the initial learning rate is 0.01. The parameters for all baselines are set as suggested in corresponding references. For each method, we conduct $5$ independent runs with random initialization, and we report the average and standard deviation of all metrics on the test set as results.

\vskip 5pt
\noindent{\bf Experimental Result:} We summarize the results of different models as follows. We first simplify our proposed model by removing the concentration feature and the attention network. The effectiveness of the concentration feature and attention network is tested independently later. The vanilla model is denoted as GBAN. The results are shown in Table \ref{tab:3}. From the table, we see GBAN has the best scores for both AUC and CTR metrics, while LibFM performs the worst.

\begin{table}[tp]
\renewcommand\arraystretch{1.5}
\centering
\caption{Comparison with baselines}
\label{tab:3}
\setlength{\tabcolsep}{5mm}{
\begin{tabular}{l c c}
\hline
Model  &AUC &CTR\\
\hline
LibFM \cite{Rendle2010Factorization} & 62.2$\pm$0.3 & 30.2$\pm$0.8\\
DeepFM \cite{Rendle2012Factorization} & 63.4$\pm$0.7 &33.5$\pm$0.5\\
YouTubeNet \cite{Covington2016Deep} & 64.8$\pm$1.0 &36.3$\pm$0.7\\		
LSTM \cite{Okura2017Embedding} & 66.8$\pm$1.3 &34.4$\pm$0.3\\		
DKN \cite{Wang2018DKN} & 66.1$\pm$1.7&34.0$\pm$0.6\\	
GBAN & \textbf{69.7}$\pm$1.1 & \textbf{39.2}$\pm$1.3\\			
\hline
\end{tabular}}
\end{table}

Based on Table \ref{tab:3}, we can draw the conclusion that deep models are effective in capturing the relations and dependencies in interactive data. This is because other deep-learning-based baselines also outperform LibFM by 1.2\% to 7.5\% on AUC and 3.3\% to 9.0\% on CTR. As for running time, we find that both DKN and LSTM take 5 hours around to train, and so does our method though we have a hybrid network of CNN and RNN. LibFM is quite different and only takes 1 hour, but its prediction is also very poor. Our method can ensure that the system is upgraded multiple times per day.

\vskip 5pt
\noindent\textbf{Test concentration feature:} To demonstrate the effectiveness of concentration feature introduced in Section \ref{subsec:ucfl}, we complement all baselines with the concentration feature and do comparisons again. The results are shown in Table \ref{tab:4}.

From Table \ref{tab:4}, we see that including the concentration feature does improve the performance for most of methods. In particular, AUC metric of DKN, LSTM, and our method increases by 1.1\%, 2.1\% and 2.9\%, respectively; and CTR metric increases by 2.9\%, 2.7\% and 3.1\%. We attribute the improvement brought by concentration feature to two reasons: (i) the coritivity helps to distinguish different users from a unitary perspective; (ii) in interactive data, the behavior sequence reflects user's concentration degree implicitly and our behavior graph can make behavior pattern clearer, which is further dug out by core and coritivity indexes.

\begin{table}[tp]
\renewcommand\arraystretch{1.5}
\centering
\caption{Comparisons with baselines by adding concentration feature}
\label{tab:4}
\setlength{\tabcolsep}{4.5mm}{
\begin{tabular}{l c c}
\hline
Model  &AUC &CTR\\
\hline
LibFM \cite{Rendle2010Factorization} +CF & 61.7$\pm$0.4 & 33.7$\pm$0.7\\
DeepFM \cite{Rendle2012Factorization} +CF & 63.6$\pm$1.2 &36.0$\pm$0.9\\
YouTubeNet \cite{Covington2016Deep} +CF & 65.5$\pm$0.9 &36.8$\pm$0.5\\		
LSTM \cite{Okura2017Embedding} +CF & 68.9$\pm$1.5 &37.1$\pm$0.7\\	
DKN \cite{Wang2018DKN} +CF & 67.2$\pm$1.1&36.9$\pm$0.7\\
GBAN +CF &  \textbf{72.6}$\pm$1.3 & \textbf{42.3}$\pm$1.4\\	
\hline
\end{tabular}}
\end{table}

\vskip 5pt
\noindent\textbf{Test attention mechanism and graph representation learning method:} We also demonstrate the effectiveness of our attention mechanism in \eqref{equ:ann} and adjusted DeepWalk method in learning nodes embeddings. In particular, we replaced adjusted DeepWalk by either standard DeepWalk \cite{Perozzi2014Deepwalk} where weights are set equally, or Node2vec \cite{Grover2016node2vec} where weights are set with $p=1$, $q=2$ (see parameter setup in \cite{Grover2016node2vec}). The results are summarized in Table \ref{tab:vgban}

\begin{table}[tp]
    \renewcommand\arraystretch{1.5}
	\centering
	\caption{Results on variants of GBAN}
	\label{tab:vgban}
	\setlength{\tabcolsep}{3.6mm}{
	\begin{tabular}{l c c}
		\hline
		Model  &AUC &CTR\\
		\hline
 		GBAN & 69.7$\pm$1.1 & 39.2$\pm$1.3\\			
		GBAN+attention & 71.6$\pm$0.7 & 40.4$\pm$0.6\\
		\hline
		GBAN with Deepwalk & 69.2$\pm$1.3 & 39.1$\pm$1.1\\		
		GBAN with Node2vec & 68.9$\pm$1.0 &38.5$\pm$0.3\\				
		\hline
		GBAN+CF & 72.6$\pm$1.3 & 42.3$\pm$1.4\\		
		GBAN+CF+attention & \textbf{73.2}$\pm$0.9 &  \textbf{43.3}$\pm$1.2\\
		\hline
	\end{tabular}}
\end{table}

From Table \ref{tab:vgban}, we see that GBAN with an attention network works better than vanilla GBAN. The AUC score increases from 69.7\% to 71.6\% and the CTR score increases from 39.2\% to 40.4\%. This is because the attention network can capture user's diverse reading interests and is candidate news-oriented. Moreover, we see that using our assigned weights in generating random walks also achieves better result than standard Deepwalk and Node2vec, due to the reflection of user's preferences and relations among nodes in assigned weights. Last, GBAN with concentration feature outperforms other variants, regardless of using the attention network or not. But our experiment shows that including both concentration feature and attention network will achieve the best score in two metrics. Thus, we conclude that all techniques used in our model are helpful to achieve a state-of-the-art performance.

\subsection{Experiment on MovieLens Dataset}\label{sec:4.4}

\vskip-0.35cm
\noindent{\bf Baseline:} We realize that a recent model called DAN \cite{Zhu2019DAN} was proposed during the review process, and has achieved good performance in recommending news. DAN introduces a timing feature extraction architecture that is related to our model. It is modified based on DKN \cite{Wang2018DKN}. In this subsection, we compete with DAN and DKN on MovieLens dataset to further demonstrate the superiority of the proposed method. However, we should mention that both DAN and DKN are based on single click behavior, so they consider a simpler problem.

Both DAN and our model adopt an attention-based LSTM to learn user representations with input news sequence. However, as explained in Remark \ref{rem:1}, their attention mechanism only processes news features without considering users differences, and our attention network is much simpler than theirs and hence can be trained faster. Moreover, DAN does not involves graph representation learning step. To study the effect of learned nodes embeddings on recommendation system and make comparison fair, we also consider modifying DKN and DAN by replacing their input feature matrices with learned embeddings. The modified DKN and DAN are denoted by DKN(*) and DAN(*). All competing methods are listed in Table \ref{tab:mbaseline}.

\begin{table}[bp]
\renewcommand\arraystretch{1.5}
\centering
\caption{Baseline models descriptions}
\label{tab:mbaseline}
\setlength{\tabcolsep}{1mm}{
\begin{tabular}{l|l}
\hline
Model  &Description \\
\hline
DKN\cite{Wang2018DKN} & Model using attention-based CNN\\			
\hline
DKN(*) & DKN with graph representation\\
\hline
DAN \cite{Zhu2019DAN} & Model using attention-based CNN and LSTM\\			
\hline
DAN(*) & DAN with graph representation\\		
\hline
\end{tabular}}
\end{table}

\vskip5pt
\noindent{\bf Parameter setup:} We let $r=9$, i.e. we use every $9$ triggered movies to predict the next one for each user. The batch size of Adam is 1000 and the initial learning rate is 0.001. For graph representation learning, we set $n=200$ (i.e. the dimension of each node embedding is 200) and all the other parameters, such as hidden layers dimensions of LSTM and CNN and setup of MMAS, are the same as the setting in Section \ref{sec:4.3}. During the training of DKN and DAN, we find that the suggested parameters of the original references lead to extremely poor performance on this dataset. We thus tune their parameters independently. The dimension of nodes embeddings and hidden layers, the number of convolution kernels of DKN and DAN are the same as GBAN. We repeat each method $5$ times with random initialization and report the average and standard deviation.

\begin{table}[tp]
\renewcommand\arraystretch{1.5}
\centering
\caption{Comparison with baselines}
\label{tab:open}
\setlength{\tabcolsep}{2mm}{
	\begin{tabular}{l c c c}
		\hline
		Model&	AUC&	AP& Kappa\\
		\hline
		DKN \cite{Wang2018DKN}&	62.30$\pm$0.17&	57.71$\pm$0.14& 24.86$\pm$0.34\\
		DKN(*)&	63.46$\pm$0.80&	59.44$\pm$1.25& 27.61$\pm$1.47\\
		\hline
		DAN \cite{Zhu2019DAN}&	61.73$\pm$0.38&	57.15$\pm$0.42& 23.55$\pm$1.06\\
		DAN(*)&	65.79$\pm$0.21&	60.59$\pm$0.21& 31.59$\pm$0.43\\
		\hline
		GBAN&	\textbf{65.96$\pm$0.22}&	\textbf{61.95$\pm$0.10}& \textbf{32.18$\pm$0.26}\\
		\hline
\end{tabular}}
\end{table}

\vskip 0.5cm
\noindent{\bf Experimental Result:}
The results are summarized in Table \ref{tab:open}. From the table, we can see that GBAN has the best score in all three metrics, AUC, AP, and Kappa. Compared with DKN and DAN, AUC of GBAN increased by 3.7\% and 4.2\%, respectively, AP increased by 4.2\% and 4.8\% and Kappa increased by 7.3\% and 8.6\%. Further, we see that DKN(*) and DAN(*) significantly improve the performance of DKN and DAN. This reveals that our constructed behavior graph with graph representation learning method can effectively exploit the complex relationships between users and movies, which further results in a more accurate recommendation. We also find in our experiment that the running time of GBAN and DKN are around 45 minutes, while DAN takes more than 1 hour. This is due to complex attention networks in DAN.

\vskip0.5cm

\section{Conclusion and future work}\label{sec:5}

In this paper, we propose a graph-based behavior-aware network (GBAN) to address the challenges for recommending interactive news. The main challenges include the analysis of multiple feedback behavior and the personalization of recommendation. We construct an interaction behavior graph for interactive news data, adopt an adjusted DeepWalk to learn node embeddings, design a graph-based CNN and an attention-based LSTM to learn news representations and users representations, and leverage tools from graph theory to use core and coritivity to measure user's concentration degree. The experimental results show that GBAN achieves the state-of-the-art performance on Tencent data and MovieLens data, outperforms recent methods DKN and DAN, and requires less running time. Further, we show in experiments that all our suggested techniques, such as concentration feature, adjustment of DeepWalk, and attention mechanism, are helpful for enhancing the performance further.

One limitation of our method is that it is not an end-to-end method. The proposed method needs to be processed in two steps. The first step is graph representation learning while the second step is the downstream behavior prediction. Two steps are trained independently. How to design a unified learning model deserves studying in the future. Also, in graph representation learning, we do not consider the heterogeneity of the graph. Designing a more appropriate representation learning method for weighted heterogeneous graphs has important applications in recommendation and related areas. {\red Our method also cannot deal with multiple edges between two nodes, which is interesting to be improved by designing a novel representation learning method.} In addition, finding other features from the behavior graph aside concentration feature to help the recommendation is also an important topic.

\vskip-0.5cm
\section*{Acknowledgements}

The work is supported by the National Key R\&D Program of China (2019YFA0706401), National Natural Science Foundation of China (61802009). Mingyuan Ma is partially supported by Beijing development institute at Peking University through award PkuPhD2019006. Sen Na is supported by Harper Dissertation Fellowship from UChicago.

\bibliographystyle{unsrt}
\bibliography{ref}


\end{document}